\documentclass{elsart}
\usepackage{latexsym,texdraw,epsfig}
\usepackage{amsfonts,amssymb,amsmath, amstext}

\newcommand{\mev}{\,\mathrm{Me\kern-0.1em V}}
\newcommand{\gev}{\,\mathrm{Ge\kern-0.1em V}}
\newcommand{\re}{\,\mathrm{{\sf I}\kern-0.1em {\sf R}}}
\newcommand{\Tr}{\mathop{\mathrm{Tr}}}
\newcommand{\hc}{\mathop{\mathrm{h.c.}}}

\begin{document}
\begin{frontmatter}
\begin{flushright}
IFUP-TH/2000-20
\end{flushright}
\title{Vortices, monopoles and confinement}
\author[Pisa]{L. Del Debbio\thanksref{lddmail}},
\author[Pisa]{A. Di Giacomo\thanksref{adgmail}},
\author[Oxford]{B. Lucini\thanksref{blmail}}

\thanks[lddmail]{ldd@df.unipi.it}
\thanks[adgmail]{digiaco@df.unipi.it}
\thanks[blmail]{lucini@thphys.ox.ac.uk}

\address[Pisa] {Dipartimento di Fisica, Universit\`a di Pisa,
and INFN Sezione di Pisa, Italy}
\address[Oxford] {Theoretical Physics, University of Oxford, UK}

\begin{abstract}
We construct the creation operator of a vortex using the me\-thods
developed for monopoles. The vacuum expectation value of this operator
is interpreted as a disorder parameter describing vortex condensation
and is studied numerically on a lattice in $SU(2)$ gauge theory. The
result is that vortices behave in the vacuum in a similar way to
monopoles.  The disorder parameter is different from zero in the
confined phase, and vanishes at the deconfining phase transition. We
discuss this behaviour in terms of symmetry. Correlation functions of
the vortex creation operator at zero temperature are also
investigated. A comparison is made with related results by other
groups.
\end{abstract}

\end{frontmatter}

\setcounter{footnote}{0}

\section{Introduction}

Recently there has been renewed interest in understanding the role of
vortices in the mechanism of colour
confinement~\cite{kovacs00,rebbi00,korthals99}.  Vortices were
originally introduced in the continuum theory as string-like
topological defects~\cite{thooft78}, and have been studied both in the
continuum and on the lattice~\cite{storiavortici}.  In the notation of
Ref.~\cite{thooft78}, a vortex creation operator $B(C)$, for a gauge
group $SU(N)$, can be associated to each closed oriented curve $C$ and
has the following commutation relation with a Wilson loop
$W(C^\prime)$, bounded by a closed curve $C^\prime$:
\begin{equation}
\label{eq:comm}
W(C^\prime) \, B(C) = B(C) \, W(C^\prime) \exp(2\pi i n/N)
\end{equation}
where $n$ is the winding number of $C^\prime$ around $C$. In $SU(2)$
such a vortex contributes a factor $(-1)$ to each Wilson loop with an 
odd linking number with the vortex line. On a lattice, the
world-sheet $M$ of the vortex can be associated to a surface on
the dual lattice. A vortex creation operator is defined by twisting
the plaquettes in the action, whose duals belong to that surface.

While creating a vortex is a well defined procedure, detecting
vortices in lattice gauge configurations is more difficult. Attempts
in this direction stemmed from~\cite{ldd97}, and results were obtained
by several groups~\cite{vort}, indicating that vortices can play a
role in confinement. However, it is not clear that this procedure
detects the vortices as defined in Eq.~\ref{eq:comm} (see
e.g.~\cite{kovacs99,jeff99}). The current picture of the role of
vortices in colour confinement is based on Eq.~\ref{eq:comm}: at low
temperature, vortices disorder the Wilson loop to produce the
well-known area law; at high temperature, vortices are suppressed and
the Wilson loop obeys a perimeter law.

In three dimensions, a conserved topological quantum number can be
associated to a vortex, and a vortex creation operator is defined as a
local complex scalar field. The dynamics of this scalar field is
described by an effective Lagrangian with a global {\it dual}\ $Z_N$
symmetry~\cite{thooft78}. The breaking of this symmetry is responsible
for the different phases of the theory. The scalar field $\phi(x_0)$
in three dimensions has a non-trivial commutation relations, analogous
to those of Eq.~\ref{eq:comm}, with any Wilson loop encircling the
point $x_0$. Operators that create such topological excitation are
known in statistical mechanics as ``disorder
operators''~\cite{kadanoff71} and the expectation value of a disorder
operator is a disorder parameter and is related to the free energy of
the associated topological excitation.

In four dimensions the situation is less clear. The vortex corresponds
to a string-like topological defect and the dual symmetry does not
emerge naturally. In this respect, there is a fundamental difference
between the vortex and the mo\-no\-po\-le picture of colour
confinement in terms of symmetry: for the monopoles, a topologically
conserved current exists, whose zero component is the generator of the
dual symmetry, and a picture based on a mechanism, namely the dual
Meissner effect \cite{dualsup}, emerges as a consequence of the
breaking of the dual symmetry~\cite{adg99}. For the vortices the
symmetry pattern is not understood. In close analogy with previous
work on abelian-projected monopoles, it is nonetheless possible to
define a vortex creation operator obeying the above commutation
relation, which can be associated to the free energy of the
vortex. Starting from this point of view, in this paper we study the
role of vortices across the deconfinement phase transition, using the
techniques developed in Refs.~\cite{ldd95,adg97,adg99} for
investigating the condensation of monopoles defined via the abelian
projection.

\newpage
A vortex creation operator $\mu$ is introduced as a disorder operator
for the $SU(2)$ lattice gauge theory. When applied to a gauge
configuration, $\mu$ creates a vortex associated to a string closing
in the $z$ direction via periodic boundary conditions (PBC). As
well as for other systems, both in statistical
mechanics~\cite{kadanoff71,jmc,dualcm} and in quantum field
theory~\cite{swieca,ldd95,adg97,adg99}, such an operator can be
constructed without any reference to the dual theory.  In
Sect.~\ref{sect:vco} we give the explicit construction of this
operator.  While our study was in progress, other papers appeared
which addressed similar problems by using quantities related to the
vortex creation operator examined in this work. The explicit
relationship between the operator studied here and other quantities in
the literature is reviewed in Sect.~\ref{sect:comp}.

Correlation functions of the vortex creation operator at zero and
finite temperature yield useful informations on the behaviour of these
topological defects in the $SU(N)$ Yang-Mills vacuum.  
As in previous studies on monopoles, it is convenient to compute
\begin{equation}
\rho = \frac{d}{d\beta} \log\langle\mu\rangle
\end{equation}
and analogous quantities for $n$-point correlators. Unlike $\mu$,
$\rho$ can be determined with good accuracy and provides all the
needed information.  The vacuum expectation value ({\em vev}) of $\mu$
proves to be a disorder parameter for the finite temperature
deconfining phase transition, as reported in
Sect.~\ref{sect:num}. Another interesting quantity to be investigated
is the correlator $\langle \mu(t_0) \mu(t_0+t)\rangle$ at zero
temperature.  

This approach can be generalised to $SU(N)$ gauge group and to vortex
loops of generic geometry, also not wrapping through PBC. Work is in
progress on these aspects.

\section{Vortex creation operator}
\label{sect:vco}

The vortex creation operator $\mu$ is constructed by the same
technique used for the monopole creation operator in previous
works~\cite{ldd95,adg97}.  The vacuum expectation value of $\mu$ is
defined as the ratio of two partition functions:
\begin{equation}
\langle \mu(t_0,x_0,y_0) \rangle = \frac{\tilde Z}{Z} = 
Z^{-1} \int [dU] e^{-\beta \,\tilde S[U]}
\end{equation}
where $Z$ is the usual partition function with the Wilson action,
\begin{equation}
S[U] = \sum_{x,\mu\nu} \Tr \left[1 - P_{\mu\nu}(x)\right]
\end{equation} 
and $\tilde S$ is obtained from $S$ by twisting a line of plaquettes
in the $0y$ plane:
\begin{equation}
\label{eq:twist}
P_{0y}(t_0, x>x_0, y_0, z) \mapsto -P_{0y}(t_0, x>x_0, y_0, z),
~\forall z
\end{equation}
The net effect of this transformation is best understood by performing
a series of changes of variables in the functional integral as
follows. The first change:
\begin{equation}
U_{y}(t_0+1, x>x_0, y_0, z) \mapsto -U_{y}(t_0+1, x>x_0, y_0, z),
~\forall z
\label{eq:change}
\end{equation}
yields
\begin{equation}
\langle \mu(t_0,x_0,y_0) \rangle = Z^{-1} \int [dU] e^{-\beta \,\tilde S^{(1)}[U]}
\end{equation}
where $\tilde S^{(1)}$ is defined by the following transformation of
the Wilson action:
\[
\left\{
\begin{array}{l}
P_{0y}(t=t_0+1, x>x_0, y_0, z) \mapsto -P_{0y}(t=t_0+1, x>x_0, y_0, z),
~\forall z \\
P_{0x} \mapsto P_{0x},~~P_{0z} \mapsto P_{0z} \\
P_{xy}(t=t_0+1,x_0,y_0,z) \mapsto
-P_{xy}(t=t_0+1,x_0,y_0,z),~\forall z \\
P_{xy}(t=t_0+1,N_s-1,y_0,z) \mapsto
-P_{xy}(t=t_0+1,N_s-1,y_0,z),~\forall z \\
P_{yz} \mapsto P_{yz},~~P_{xz} \mapsto P_{xz} \\
\end{array}
\right.
\]

\begin{figure}[ht]
\begin{center}
\epsfig{figure=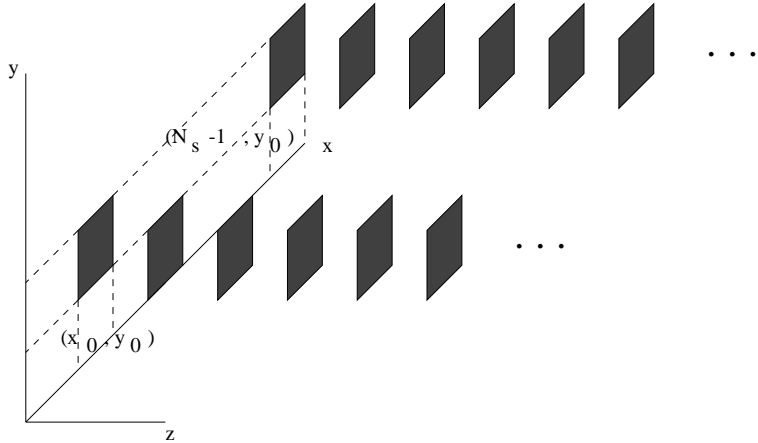, width=10cm} 
\end{center}
\caption{Set of plaquettes changing sign in the hyper-plane $t_0+1$ after
the first change of variables described in Eq.~\ref{eq:change}. The
plaquettes in the plane $xy$ correspond to the location of the vortex
lines at $(x_0,y_0)$, and at $(N_s-1,y_0)$ due to PBC in the $x$
direction.}
\label{fig:def}
\end{figure}
The change of sign in the $P_{0y}$ plaquettes introduced at $t=t_0$ in
Eq.~\ref{eq:twist} has been shifted at $t=t_0+1$ and two vortices have
been created in the $(x,y)$ plane for $t=t_0+1$ at $(x_0,y_0)$ and
$(N_s-1,y_0)$. It is worthwhile to remark that the second vortex at
$(N_s-1,y_0)$ is due to the fact that we are working in a finite volume
with PBC in the $x$ direction. In an infinite volume, or with free BC,
the vortex at $(x_0,y_0)$ would have been created alone. In the $z$
direction, the vortices either extend to infinity (for an infinite
lattice), or they form a closed loop due to PBC. The set of $xy$
plaquettes changing sign after this first change of variables is shown
in Fig.~\ref{fig:def}.

The same change of variables can be iterated at successive times,
yielding, after $n$ iterations:
\begin{equation}
\langle \mu(t_0,x_0,y_0) \rangle = Z^{-1} \int [dU] e^{-\beta \,\tilde S^{(n)}[U]}
\end{equation}
and $\tilde S^{(n)}$ is obtained from the Wilson action via:
\[
\left\{
\begin{array}{l}
P_{0y}(t=t_0+n, x>x_0, y_0, z) \mapsto -P_{0y}(t=t_0+n, x>x_0, y_0, z),
~~~~~\forall z \\
P_{xy}(t=t_0+1,x_0,y_0,z) \mapsto
-P_{xy}(t=t_0+1,x_0,y_0,z),~\forall z \\
P_{xy}(t=t_0+1,N_s-1,y_0,z) \mapsto
-P_{xy}(t=t_0+1,N_s-1,y_0,z),~\forall z \\
\hfill \vdots \hfill \\
P_{xy}(t=t_0+n,x_0,y_0,z) \mapsto
-P_{xy}(t=t_0+n,x_0,y_0,z),~\forall z \\
P_{xy}(t=t_0+n,N_s-1,y_0,z) \mapsto
-P_{xy}(t=t_0+n,N_s-1,y_0,z),~\forall z \\
\end{array}
\right.
\]
Such a configuration corresponds to the propagation of the two
vortices, from time $t=t_0$ to time $t=t_0+n$.

Correlations of $\mu$ operators at different times $\{t_1, \ldots,
t_n\}$ are defined by repeating the transformation in
Eq.~\ref{eq:twist} for each $t_i$.
For a two-point correlation:
\begin{equation}
\label{eq:corr}
\Gamma(t) = \langle \mu(t_0,x_0,y_0) \,\mu(t_0+t,x_0,y_0) \rangle
\end{equation}
the change in the $P_{0y}$ plaquettes is
reabsorbed when $n=t$ and the correlator describes the propagation of
a pair of vortices from the time $t_0$ to the time $t_0+t$. 

We remark here that these prescriptions create vortex lines that are
closed in space, due to PBC in the $z$-direction, and propagate in
time. 

At large values of $t$, $\Gamma$ decreases exponentially to an
asymptotic value, which, by cluster property, is $\langle \mu
\rangle^2$, the square of the disorder parameter:
\begin{equation}
\Gamma(t) \simeq A e^{- m t} + \langle \mu \rangle^2
\label{eq:time-b}
\end{equation}
At $T=0$, $\langle\mu\rangle$ can be extracted from the large-$t$
value of the correlator, according to Eq.~\ref{eq:time-b}.

At finite temperature, there is no propagation in time, and a direct
measurement of $\langle\mu\rangle$ is needed by use of a single
operator. In order to have a consistent implementation of the changes
of variables described above, $C^*$ boundary conditions are
needed in the time direction, as explained in~\cite{adg97}. 

Closed vortex lines propagating in time can be created in a field
configuration with the same prescription, but with a different
modification of the action. For instance, using for the modified action:
\begin{equation}
P_{0y}(t_0, x_0<x<x_1, y_0, z_0<z<z_1) \mapsto -P_{0y}(t_0, x_0<x<x_1, y_0, z_0<z<z_1)
\label{eq:def_clo}
\end{equation}
gives a vortex line encircling the rectangle in the $xz$ plane:
\begin{equation}
R = \left\{ \left( x, y, z \right) : x_0<x<x_1, y=y_0, z_0<z<z_1 \right\}
\end{equation}
The same definition is used in a recent
publication~\cite{cheluvaraja00}.

\section{Comparison with related works}
\label{sect:comp}

Our disorder parameter is strictly related to observables introduced
by other authors~\cite{kovacs00,rebbi00,korthals99}. The detailed
comparison is as follows.

The paper~\cite{rebbi00} presents a computation of the free energy of
a vortex pair. The latter is defined as:
\begin{equation}
\label{eq:rebbi_def}
F = -T \log \frac{Z(\beta,-\beta)}{Z(\beta,\beta)}
\end{equation}
where 
\begin{eqnarray}
Z(\beta,\beta^\prime) &=& \int [dU] e^{-S(\beta,\beta^\prime)}
\nonumber \\
S(\beta,\beta^\prime) &=&  \frac12 \left(\beta \sum_{P\not\in M} \Tr P
+ \beta^\prime \sum_{P\in M} \Tr P \right)
\nonumber
\end{eqnarray}
and $T=(N_t a)^{-1}$ is the temperature of the system. $M$ indicates a
set of plaquettes with modified coupling $\beta^\prime=-\beta$, which
can be seen as the plaquettes transversed by the vortex string. The
following two cases are examined:
\begin{itemize}
\item a vortex solution is placed at $(x_0,y_0,z_0,0)$ and an
anti-vortex at $(x_0,y_0,z_0+d,0)$, with a straight string of twisted
plaquettes connecting them;
\item a single vortex is placed in the middle of the lattice and the
string extends to the boundary in the $z$-direction, where free
boundary conditions are used.
\end{itemize}
In both cases the modification of the action is done in all
time-slices.

Using the notation introduced in the previous section:
\begin{equation}
\frac{Z(\beta,-\beta)}{Z(\beta,\beta)} = Z^{-1} \int [dU]
e^{-S^\prime[U]}
\end{equation}
and now $S^\prime$ is obtained from $S$ by:
\[
\left\{
\begin{array}{l}
P_{xy}(t,x_0,y_0,z=z_0+1) \mapsto
-P_{xy}(t,x_0,y_0,z=z_0+1),~\forall t \\
\hfill \vdots \hfill \\
P_{xy}(t,x_0,y_0,z=z_0+d) \mapsto
-P_{xy}(t,x_0,y_0,z=z_0+d),~\forall t \\
\end{array}
\right.
\]
With a $t \rightleftharpoons z$ relabelling of the axes, the
correlator $\Gamma$ defined in this paper by Eq.~\ref{eq:corr} yields
the free energy of {\it two}\ vortex pairs as defined in~\cite{rebbi00} at
distance $N_s/2$.  

At $T=0$ we use symmetric lattices ($N_s=N_t$) and measure the time
correlator of two $\mu$ operators at $x_0=y_0=N_s/2$, which is in fact
the correlator of two pairs, when the effect of periodic boundary
conditions are taken into account. At non-zero temperature, we use a
single vortex operator, which actually introduces two vortices, again
due to PBC. The extra vortices that are created by PBC are $N_s/2$
lattice spacings away from their partners. If $N_s \gg \xi$, where
$\xi$ is the correlation length, then our disorder parameter $\langle
\mu\rangle$ is the square of $\exp\left(-F/T\right)$. 

\begin{figure}[htp]
\begin{center}
\epsfig{figure=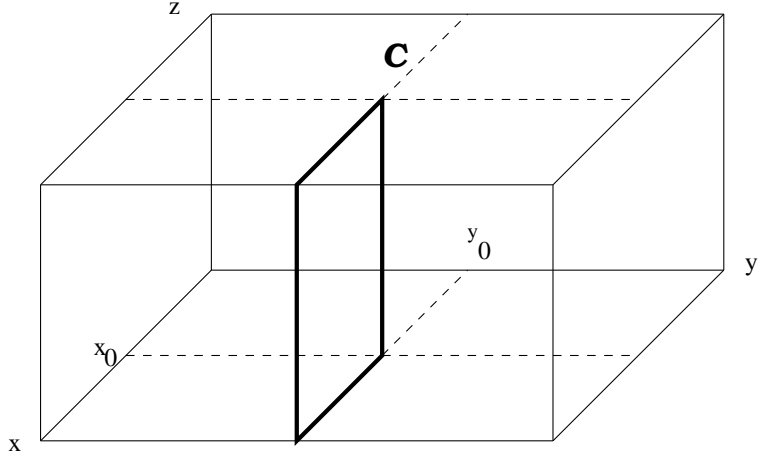, width=10cm} 
\end{center}
\caption{The operator $\mu$ introduced in sect.~\ref{sect:vco}
corresponds to $V(C)$ for the curve $C$ depicted here. The box
represents a time-slice of the whole lattice.}
\label{fig:curve}
\end{figure}

The authors of Ref.~\cite{korthals99} are concerned with the behaviour
of the 't~Hooft loop~\cite{thooft78} in hot QCD, and relate the dual
string tension to the wall tension of $Z_N$ domain walls. For $SU(2)$,
the lattice definition of the loop operator is given
by~\cite{korthals81}:
\begin{equation}
\label{eq:VC}
V[C] = \exp \left\{ \beta \sum_{x\in S} \left( \Tr P_{zt}(x) +
\hc \right) \right\}
\end{equation}
where $S$ is the surface in the $xy$ plane bounded by the curve
$C$. The vev of the operator defined in Eq.~\ref{eq:VC} is again the
ratio of two partition functions: one with a modified action divided
by the (standard) Wilson action. The modified action is obtained by
changing the sign of the plaquettes in $S$. The definition given in
sect.~\ref{sect:vco} coincides exactly with Eq.~\ref{eq:VC}, for the
curve $C$ depicted in Fig.~\ref{fig:curve}, after a relabelling of the
axes $y \rightleftharpoons z$. It is easy to realise in this
formulation that our vortex creation operator is precisely a 't~Hooft
loop operator: every Wilson loop with non-trivial linking to the curve
$C$ receives a $(-1)$ contribution from the vortex.

Finally, the authors of Ref.~\cite{kovacs00} compute the free energy
of a closed $SU(2)$ vortex, which is defined from the logarithm of
the ratio of two partition functions:
\begin{equation}
\exp\left\{-F(\tau_{\mu\nu}) \right\} = \frac{Z(\tau_{\mu\nu})}{Z}
\end{equation}
$Z(\tau_{\mu\nu})$ is defined as usual by multiplying a given
(co-closed) set ${\mathcal V}_{\mu\nu}$ of plaquettes in the action with
an element $\tau_{\mu\nu}$ of the center of the gauge group. For
$SU(2)$, the only non-trivial possibility is $\tau_{\mu\nu}=-1$. 
For the set of plaquettes:
\[
{\mathcal V}_{\mu\nu} = \left\{P_{\mu\nu}(t,x,y,z) : (\mu,\nu) = (0,2), t=t_0,
x_0<x<x_1, y=y_0, z_0<z<z_1 \right\}
\]
the definition in~\cite{kovacs00} coincides with the one in
Eq.~\ref{eq:def_clo}.

\section{Numerical results}
\label{sect:num}

In this paper, we present data only for open vortex lines wrapping in
the $z$ direction by PBC. Work is in progress to study closed vortices
as done in~\cite{cheluvaraja00}.

Due to the exponential in its definition, $\mu$ has large
fluctuations, which make a direct measurement a challenging
task. Instead, as in previous studies~\cite{ldd95,adg99}, we define
\begin{equation}
\rho = \frac{\mbox{d}}{\mbox{d} \beta} \log \langle \mu \left( t_0, x_0, y_0
\right) \rangle = \langle S \rangle _S - \langle \tilde{S} \rangle _{\tilde{S}}
\end{equation}
Being the difference of the expectation values of two actions, $\rho$
can be easily computed, and proves to yield all the relevant
information.  At finite temperature, $\rho$ is expected to have a
sharp negative peak in the critical region~\cite{ldd95,adg97,adg99},
if $\langle \mu \left( t_0, x_0, y_0) \right \rangle$ is related to
the deconfinement phase transition.

\begin{figure}[htp]
\begin{center}
\epsfig{figure=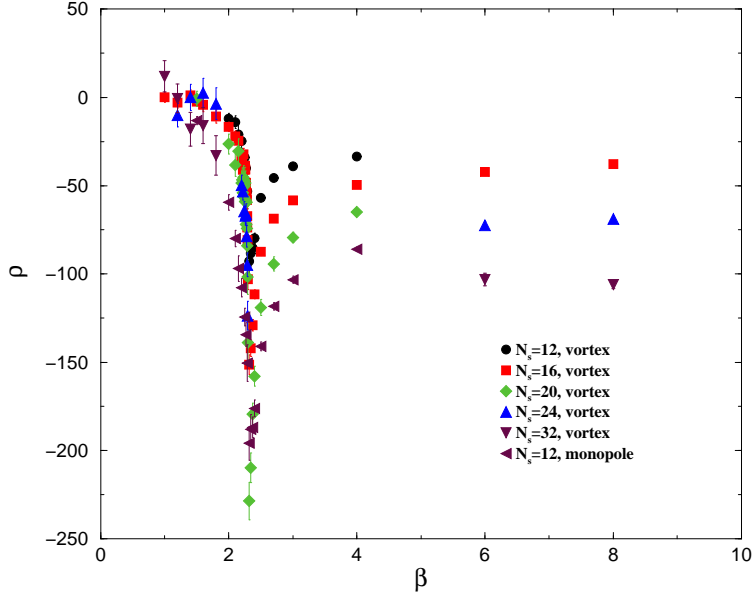, angle=270, width=10cm} 
\end{center}
\caption{Comparison of $\rho$ defined with the vortex
and the monopole creation operator.} 
\label{fig:finitet1}
\end{figure}

Our results for $\rho$ are displayed in Fig.~\ref{fig:finitet1}, for
several lattice sizes. For comparison, we also report a plot of the
corresponding quantity for the {\em vev} of a monopole creation
operator on a $12^4 \times 4$ lattice \cite{adg99}. Already at a
qualitative level, it is clear that, at weak coupling, $\rho$ is
negative, with its absolute value increasing with the volume. This
behaviour implies that $\langle \mu \rangle$ vanishes for large
$\beta$ in the thermodynamical limit. At low $\beta$, $\rho$ is
compatible with 0 for all the volumes considered in this study, which
means that $\langle \mu \rangle$ has a non-zero value in the infinite
volume limit in the confined phase. The presence of the negative peak
connecting the confined and deconfined phases suggests that vortices
play a role at the deconfinement transition. This behaviour is in
agreement with the results in~\cite{kovacs00,rebbi00}. As in the case
of monopoles~\cite{adg99}, we can perform a finite size scaling
analysis of the data presented above. In neighborhood of the critical
coupling $\beta_c$, if $\langle \mu \rangle$ is a disorder parameter
for the deconfining phase transition, in the infinite volume limit we
have
\begin{eqnarray} 
\langle \mu \rangle \propto \left(\beta_c - \beta \right)^{\delta}  
\end{eqnarray}
being $\delta$ the corresponding critical index. In a finite volume,
the previous equation is replaced by
\begin{eqnarray}
\label{musca1}
\langle \mu \rangle = \left(\beta_c - \beta \right)^{\delta} \Phi(N_s/\xi)
\end{eqnarray}
where $\xi$ is the correlation length and $\Phi$ is a function of
the ratio $N_s/\xi$. Since
\begin{eqnarray}
\xi \propto \left( \beta_c - \beta \right)^{- \nu}
\end{eqnarray}
Eq.~(\ref{musca1}) can be written as
\begin{eqnarray}
\label{musca2}
\langle \mu \rangle = L^{-\delta/\nu} \tilde{\Phi}(L^{1/\nu}(\beta_c - \beta))
\end{eqnarray}
being $\nu$ the critical index of the correlation length.

Eq.~(\ref{musca2}) implies
\begin{eqnarray}
\frac{\rho}{L^{1/\nu}} = f\left(L^{1/\nu}
\left(\beta_c - \beta\right)\right)
\end{eqnarray}
i.e. the ratio $\rho/L^{1/\nu}$ is an universal function of the scaling
variable 
\[
x=L^{1/\nu}\left(\beta_c - \beta\right)
\]

\begin{figure}[htp]
\begin{center}
\epsfig{figure=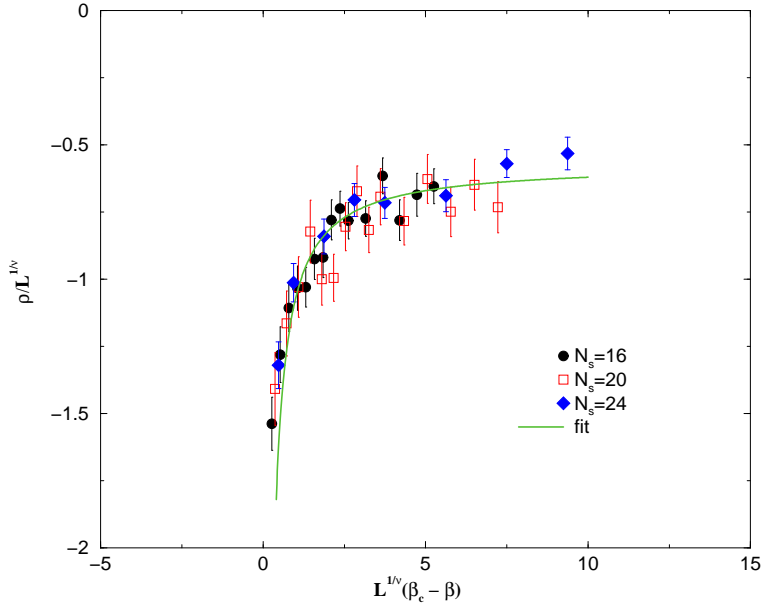, angle=270, width=10cm} 
\end{center}
\caption{Plot of rescaled data.} 
\label{figscaling}
\end{figure}

By guessing that \cite{adg99}
\begin{eqnarray}
\frac{\rho}{L^{1/\nu}} = - \frac{\delta}{x} + c
\end{eqnarray}
it is possible to extract from our data the critical exponents
$\delta$ and $\nu$ and the critical coupling $\beta_c$. We perform
three different fits: a fit with three parameters, a fit to $\delta$
and $\nu$ at fixed $\beta_c$, using the value in~\cite{karsch93}, and
again a two parameter fit to $\delta$ and $\nu$ at fixed $\beta_c$,
using the value obtained in the three parameter fit. The error is
estimated from the variation in the fitted values when using the
di\-fferent methods described above and when the points corresponding
to smaller correlation lengths are excluded. Our results are:
\begin{eqnarray}
\beta_c &=& 2.30 (1) \nonumber \\
\delta  &=& 0.5  (1) \nonumber \\
\nu     &=& 0.7  (1) \nonumber
\end{eqnarray}
The values of $\beta_c$ and $\nu$ are in good agreement with the
values obtained in \cite{adg99}, while the value of $\delta$ varies by
two standard deviations, when compared to previous results. 

Fig.~\ref{figscaling} shows how well the scaling relation is obeyed
with the values of $\beta_c$ and $\nu$ from the above fit.

Assuming that in the weak coupling limit all link variables are close
to unity, and remembering that the twisted action is obtained by
flipping the sign of $2 \times N_t \times L$ plaquettes, a naive prediction would
yield:
\begin{equation}
\rho \simeq - 16 L + \mathrm{const}
\label{eq:naivept}
\end{equation}
This is a justification for the linear dependence used in the ansatz
above. However, the actual value of the coefficient $a$ does not need
to be the one in Eq.~\ref{eq:naivept}, since the configuration with
all links set to the identity is not the true minimum of the modified
action which enters the definition of $\rho$.

To check this behaviour, we have measured $\rho$ for different volumes
and very large values of the coupling. In this region, we expect the
data to be independent of $\beta$. Within the errors the data do lie
on a straight line as predicted by Eq.~\ref{eq:naivept}. The fact that
the slope is negative ensures that the disorder operator
vanishes in the weak coupling phase in the thermodynamic limit.

\begin{figure}[htp]
\begin{center}
\epsfig{figure=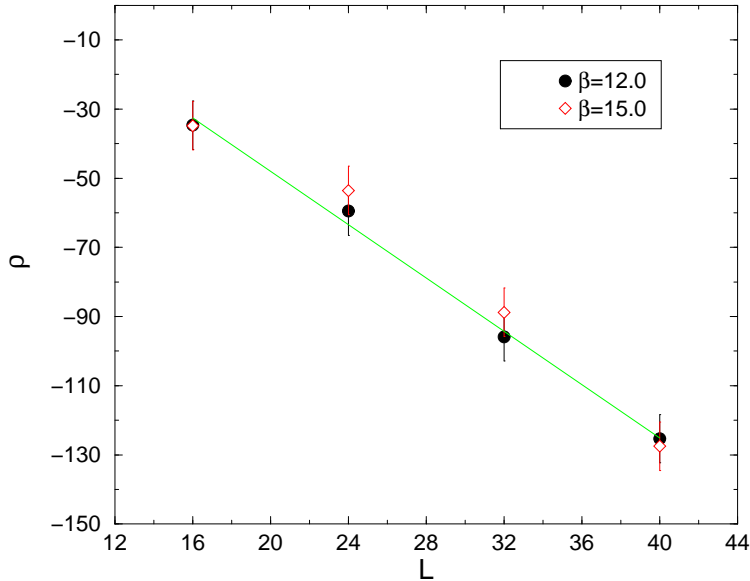, width=10cm}
\end{center}
\caption{$\rho$ vs. $L$ for large values of $\beta$. The solid line
corresponds to a linear fit.}
\label{fig:wc_vs_L}
\end{figure}

\begin{figure}[htp]
\begin{center}
\epsfig{figure=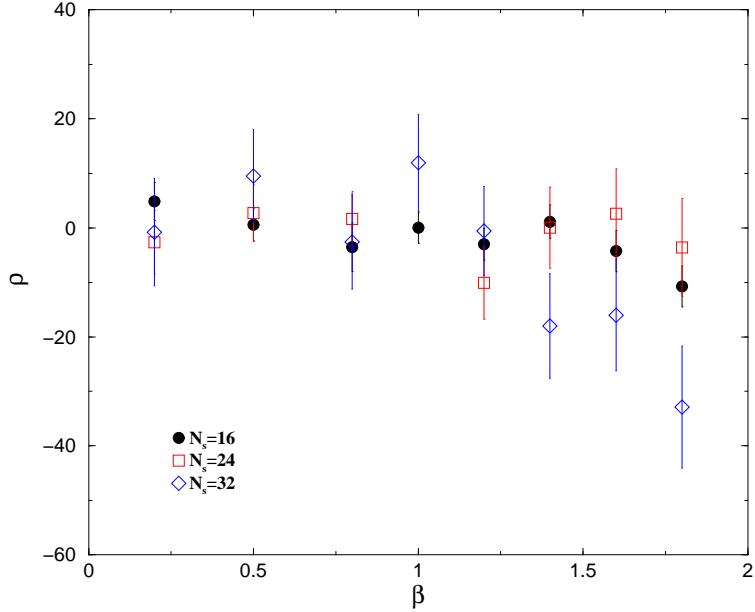, angle=270, width=10cm} 
\end{center}
\caption{$\rho$ vs. $\beta$ for different spatial volumes.}
\label{fig:sc}
\end{figure}

The behaviour of $\rho$ at low $\beta$'s is displayed in
Fig.~\ref{fig:sc}: the value at low $\beta$ remains bounded from below
and consistent with zero for increasing volumes, which guarantees that
$\rho$ does not vanish in the thermodynamic limit above the phase
transition.

Another quantity that can give information on the behaviour of center
vortices in the vacuum of SU(2) is the correlation function of two $\mu$ 
operators as a function of distance at zero temperature. When relabelling
the axes, as discussed in the previous Section, this quantity
is identical to the free energy recently discussed in \cite{rebbi00}.

Again, our numerical computation is performed in terms of $\rho_2$,
now defined as
\begin{equation}
\rho_2(t) = \frac{\mbox{d}}{\mbox{d} \beta} \log \langle \mu \left( t_0, x_0, y_0
\right) \mu \left( t_0 + t, x_0, y_0 \right) \rangle 
\end{equation}
which is the difference between two actions. Since the usual Wilson
contribution is $t$-independent, the whole dependence on $t$ is brought by
$\tilde{S}$. 

\begin{figure}[htp]
\begin{center}
\epsfig{figure=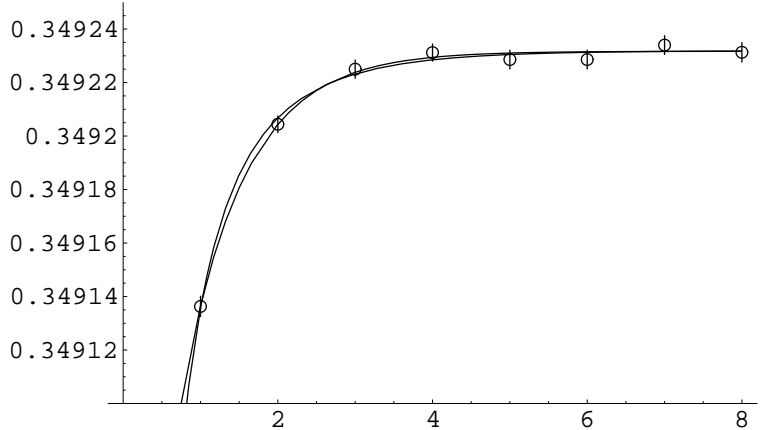, width=10cm} 
\end{center}
\caption{The modified action $\tilde{S}$ per unit of volume as a
function of the distance between the vortex and the antivortex on a
lattice $16^4$ at $\beta = 2.5$.}
\label{fig:corr1}
\end{figure}

In Fig.~\ref{fig:corr1} we display the behaviour of $\tilde{S}$ on a $16^4$
lattice at $\beta = 2.5$, where the system is confined and scaling
is supposed to work.

The $t$ dependence of the correlation is fitted using two different
fitting functions, both depending on three parameters $(a,b,m)$:
\begin{eqnarray}
\rho_2(t) &=& a \frac{e^{-m t}}{t} + b \nonumber \\
\rho_2(t) &=& a e^{-m t} + b \nonumber 
\end{eqnarray}
these two fits will be named respectively fit I and fit II in what
follows.  The functional dependence in fit I describes a Yukawa
potential between vortices, as studied in Ref.~\cite{rebbi00}. In both
cases, $b$ is the asymptotic value of $\rho_2$ and is expected to be
different from zero if vortices are condensed.  Both ansatz fit the
data relatively well, despite the fact that the masses obtained are
quite different. The values of $\chi^2$ give an indication that fit II
could be a better description of the data, although it is impossible
to draw any robust conclusion given that it is very difficult to
disentangle the logarithmic correction in $t$ which differentiate fit
I from fit II. The outcome of the fits is summarised in
Table~\ref{tab:fit_res} and the two fitting curves are superimposed on
the data in Fig.~\ref{fig:corr1}. The value of the constant $b$ is
determined quite accurately due to the plateau in the data points at
large $t$. It is interesting to remark that $a$ turns out to be
negative in both cases up to 90\% CL, while the error on the fitted
mass is very large. This is mainly due to the fact that large
variations in $m$ can be reabsorbed by variations in $a$ in the range
where data points are available. 

\begin{table}[htp]
\begin{center}
\begin{tabular}{lrrrr}
\hline 
fit type & $a$      & $b$     & $m$   & $\chi^2$/d.o.f. \\
\hline
fit I    & -0.00018 & 0.34923 & 0.650 & 2.6      \\
fit II   & -0.00035 & 0.34923 & 1.302 & 1.8      \\
\hline
\end{tabular}
\end{center}
\caption{Fit results for the correlation of vortex operators}
\label{tab:fit_res}
\end{table}

We notice from fit II that the correlation length is around 1.3
lattice spacings. This justifies the view of the pair of vortices (the
one at the center of the lattice and the one at the border produced by
PBC) as independent.

\newpage

\section{Discussion and future outlooks}
\label{sect:conc}

In previous works~\cite{adg99}, we produced evidence of dual
superconductivity of the QCD vacuum, supporting the mechanism of
Refs.~\cite{dualsup}. A peculiar feature of this phenomenon is that
monopoles defined by different abelian projections all
condense~\cite{adg99}. This possibility was suggested in
Ref.~\cite{thooft81} and lattice data support it, showing that
confinement is related to condensation of magnetic charges defined by
a few abelian projections. 

If all or a large class of abelian projections are equivalent, there
are infinitely many physically equivalent disorder symmetries. We
already observed that this fact is not inconsistent, but suggests that
maybe a more fundamental dual symmetry pattern exists, which can
manifest itself as condensation of all these magnetic charges~\cite{adg99a}. 

Now it is found that also vortices show condensation. We were not able
to associate a dual conserved topological quantity to vortices in 3+1
dimensions, contrary to what happens in 2+1 dimensions. However we
consider what is observed here an important information on the way to
understand the features of the dual description.

We are extending the analysis to the vortices of the $SU(3)$ gauge
theory and to some aspects of closed vortex lines.  We are also trying
to understand what happens in the presence of quarks. Preliminary data
on condensation of monopoles, as defined by the abelian projections,
show that the presence of quarks does not affect the dual
superconductor picture. This is in agreement with the idea that
the gross features of the QCD vacuum, including the mechanism of
confinement, are determined already at $N_c=\infty$~\cite{nlimit}.

We are now trying to understand how to extend to full QCD the analysis
of vortices. 

\bigskip
\noindent
{\bf Acknowledgements} We thank M. D'Elia, G. Paffuti and M. Teper for
enlightening discussions. Financial support by the EC TMR Pro\-gram
ERB\-FMRX-CT97-0122 and by MURST is acknowledged. BL is funded by PPARC
under Grant PPA/G/0/1998/00567.

\end{document}